\documentclass[iop]{emulateapj}   

\usepackage{ulem}

\usepackage{epsfig}
\usepackage{color}
\usepackage{natbib}
\newcommand{\fesc}{\ifmmode{f_{\rm esc}}\else{$f_{\rm esc}$}\fi}
\newcommand{\fescs}{\ifmmode{f_{\rm esc}^\star}\else{$f_{\rm esc}^\star$}\fi}
\newcommand{\kms}{\ifmmode{{\;\rm km~s^{-1}}}\else{km~s$^{-1}$}\fi}
\newcommand{\fgas}{\ifmmode{{f_{\rm gas}}}\else{$f_{\rm gas}$}\fi}
\newcommand{\cubecm}{\ifmmode{{\rm cm^{-3}}}\else{cm$^{-3}$}\fi}
\newcommand{\ztwo}{\ifmmode{{\rm [Z_2/H]}}\else{[Z$_2$/H]}\fi}
\newcommand{\zthree}{\ifmmode{{\rm [Z_3/H]}}\else{[Z$_3$/H]}\fi}
\newcommand{\lsim}{\lower0.3em\hbox{$\,\buildrel <\over\sim\,$}}
\newcommand{\gsim}{\lower0.3em\hbox{$\,\buildrel >\over\sim\,$}}

\newcommand{\emis}{erg s$^{-1}$ cm$^{-2}$ Hz$^{-1}$ sr$^{-1}$}

\newcommand{\eavg}{\ifmmode{\langle E_\gamma \rangle}\else{$\langle E_\gamma \rangle$}\fi}

\newcommand{\Ms}{\ifmmode{\textrm{M}_\odot}\else{$M_\odot$}\fi}
\newcommand{\vrms}{\ifmmode{v_{\rm rms}}\else{$v_{\rm rms}$}\fi}

\newcommand{\hh}{H$_2$}

\newcommand{\tvir}{\ifmmode{T_{\rm{vir}}}\else{$T_{\rm{vir}}$}\fi}
\newcommand{\mvir}{\ifmmode{M_{\rm{vir}}}\else{$M_{\rm{vir}}$}\fi}
\newcommand{\rvir}{\ifmmode{r_{\rm{vir}}}\else{$r_{\rm{vir}}$}\fi}

\newcommand{\jj}{\ifmmode{J_{21}}\else{$J_{21}$}\fi}
\newcommand{\flw}{\ifmmode{F_{LW}}\else{$F_{LW}$}\fi}
\newcommand{\kph}{\ifmmode{k_{\rm ph}}\else{$k_{\rm ph}$}\fi}

\newcommand{\zsun}{\ifmmode{\rm\,Z_\odot}\else{$\rm\,Z_\odot$}\fi}

\begin{document}

\shorttitle{POP III STARS AND REMNANTS}
\shortauthors{XU ET AL.}

\title{Population III Stars and Remnants in High Redshift Galaxies }


\author{Hao Xu\altaffilmark{1}, 
John H. Wise\altaffilmark{2},    
and Michael L. Norman\altaffilmark{1}}
\altaffiltext{1}{Center for Astrophysics and Space Sciences,
  University of California, San Diego, 9500 Gilman Drive, La Jolla, CA
  92093; hxu@ucsd.edu, mlnorman@ucsd.edu}
\altaffiltext{2}{Center for Relativistic Astrophysics, School of
  Physics, Georgia Institute of Technology, 837 State Street, Atlanta,
  GA 30332; jwise@gatech.edu}

\begin{abstract}

  Recent simulations of Population III star formation have suggested
  that some fraction form in binary systems, in addition to having a
  characteristic mass of tens of solar masses.  The deaths of
  metal-free stars result in the initial chemical enrichment of the
  universe and the production of the first stellar-mass black holes.
  Here we present a cosmological adaptive mesh refinement simulation
  of an overdense region that forms a few $10^9\; \Ms$ dark matter halos
  and over 13,000 Population III stars by redshift 15.  We find that
  most halos do not form Population III stars until they reach $M_{\rm
    vir} \sim 10^7\; \Ms$ because this biased region is quickly enriched
  from both Population III and galaxies, which also produce high
  levels of ultraviolet radiation that suppress H$_2$ formation.
  Nevertheless, Population III stars continue to form, albeit in more
  massive halos, at a rate of $\sim 10^{-4} \Ms \, \textrm{yr}^{-1} \,
  \textrm{Mpc}^{-3}$ at redshift 15.  The most massive starless halo
  has a mass of $7 \times 10^7\; \Ms$, which could host massive black
  hole formation through the direct gaseous collapse scenario.  We
  show that the multiplicity of the Population III remnants grows with
  halo mass above $10^8 \; \Ms$, culminating in 50 remnants located in
  $10^9 \; \Ms$ halos on average.  This has implications that high mass
  X-ray binaries and intermediate mass black holes that originate from
  metal-free stars may be abundant in high-redshift galaxies.

\end{abstract}

\section{Introduction}
\label{introduction}

The first generation stars (Population III) form from metal-free gas
in dark matter halos with M $\sim$ 10$^{6}$ M$_{\odot}$ and have a
large characteristic mass \citep[e.g.][]{Abel02, Bromm02, OShea07a,
  Turk09, Greif12_P3Cluster}. Due to their high mass, they have short
lifetimes \citep{Schaerer02} and may go supernova, and enrich their
surrounding intergalactic medium (IGM).  For these metal-free stars,
Type II supernovae (SNe) happen for an initial mass between 10 and 40
M$_{\odot}$, and much more energetic pair-instability SNe (PISNe)
might occur in stars between 140 and 260 M$_{\odot}$ \citet{Heger03}.
Once the metallicity passes some critical metallicity, $\sim$
10$^{-6}$ Z$_{\odot}$ if dust cooling is efficient
\citep{Omukai05,Schneider06, Clark08} or $\sim$ 10$^{-3.5}$
Z$_{\odot}$ otherwise \citep{Bromm01, Smith09}, the gas can cool
rapidly and lower its Jeans mass.  These metal-enriched (Population
II) stars have a lower characteristic mass scale and most likely have
an initial mass function (IMF) that resembles the present-day one.
However at high redshift, heating from the cosmic microwave background
(CMB) may limit the radiative cooling, and thus increasing the Jeans
mass, resulting in an IMF that also favors massive star formation
\citep{Larson05, Smith09}.

The transition from Pop III to Pop II star formation is solely
dependent on the metal enrichment from the Pop III SN remnants in the
future star forming halos. Metal enrichment involves complex
interactions between SNe blastwaves, the intergalactic medium (IGM),
halo mergers, and cosmological accretion.  This topic has been
extensively studied with semi-analytic models \citep{Scannapieco03,
  Yoshida04, Tumlinson06, Salvadori07, Komiya10}, post-processing of
numerical simulations \citep{Karlsson08, Trenti09}, and direct
numerical simulations \citep{Tornatore07, Ricotti08, Maio10_Pop32,
  Wise12a, Muratov12}. For example, \citet{Trenti09} suggested that
Pop III stars may still form at the late epoch of $z = 6$ in the under
dense regions of the universe by post-processing of cosmological
simulations with blast wave models. \citet{Muratov12} also showed that
Pop III stars continue to form until $z = 6$ using direct cosmological
simulations.

In addition to the metal enrichment from Pop III stars, heating and ionizing
effects from their radiation are crucial to modeling early structure
formation of the Universe. \citep{Haiman00}.  Lyman-Werner photons
from Pop III stars photodissociate \hh~by the Solomon process and suppress the
formation of Pop III stars in low-mass halos. The higher energy UV
radiation from Pop III stars then affects the subsequent structure
formation through heating and ionizing the surrounding IGM 
\citep{Machacek01,Yoshida03,Wise07_UVB,OShea08}. Pop III
more massive than 260 M$_\odot$ or less massive than 140 M$_\odot$ may
direct collapse to form black holes \citet[BHs;][]{Heger03}. Accretion
onto these massive Pop III BHs is a feasible way to form $z>7$ quasars
\citep{Johnson12} and is also an important source of X-ray radiation
in high redshifts.  X-rays from the accretion onto Pop III BHs
\citep{Alvarez09} or from Pop III binaries \citep{Turk09, Stacy10,
  Stacy12} may preheat and pre-ionize a large volume of the IGM
\citep{Ricotti04, Mesinger13}.

\begin{figure*}
\begin{center}
\epsfig{file=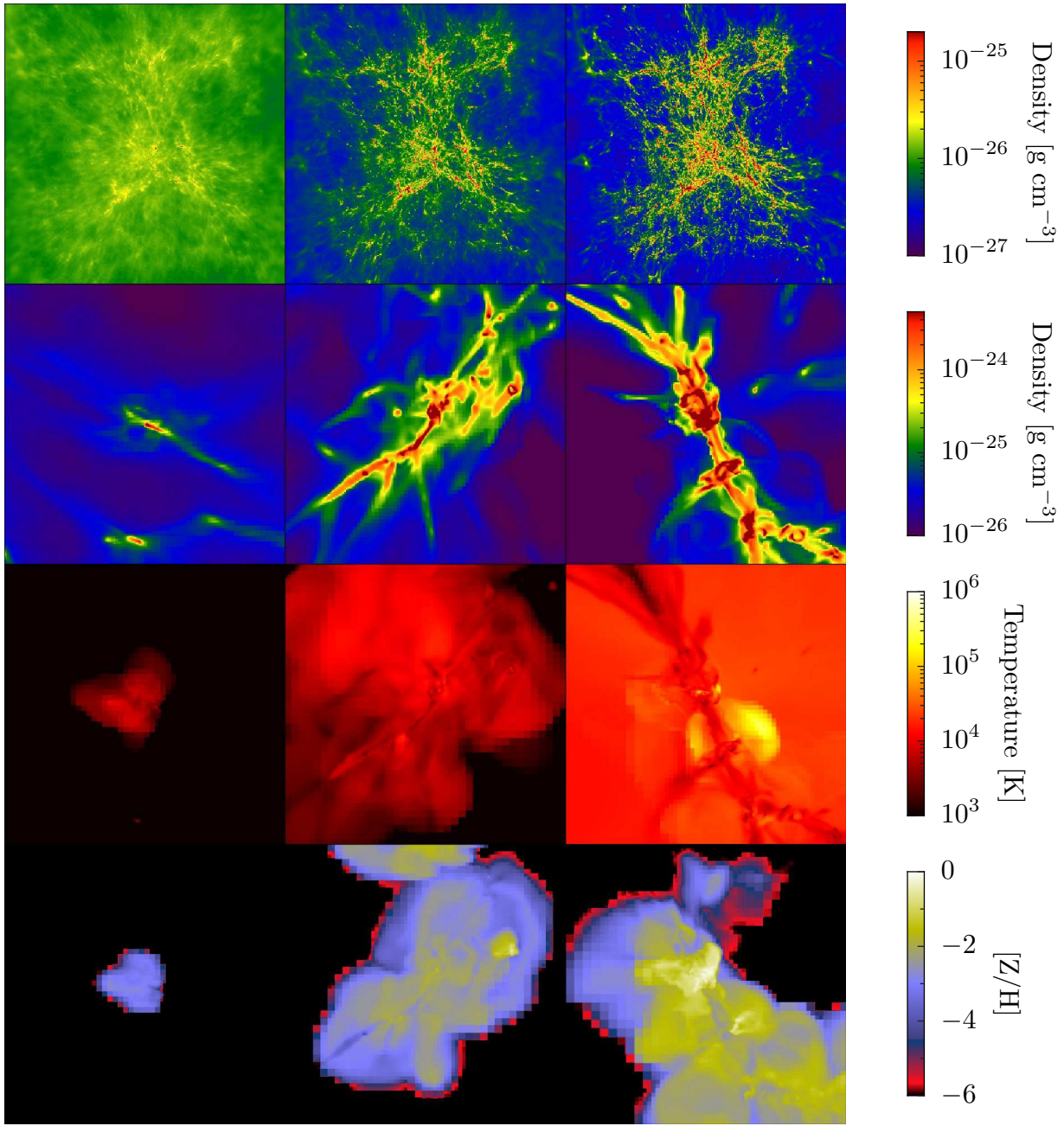,width=0.9\textwidth}
\end{center}

\caption{Snapshots of the refined regions and the most massive halos
  at redshifts 25 (left), 17.91 (middle) and 15 (right). The images
  are the density-weighted projections of baryon density in cubic
  volumes 6.6 comoving Mpc on a side (first row), enclosing the
  refined regions, the density-weighted projections of baryon density
  (second row), temperature (third row) and metallicity (fourth row)
  in cubic volumes 10.0 proper kpc on a side, enclosing the most
  massive halos. 
  \label{fig:snapshots}}
\end{figure*}
 
The impact of Pop III feedback on cosmic evolution is dependent on the
properties of their host halos, especially their masses
\citep{Whalen08, Muratov12}. The sizes of the host halos determine the
distance the metal from Pop III SNe reach with their blast waves,
and the escape faction of their UV radiation. This makes a detailed
study of the Pop III star and remnant distribution over a wide range
of high redshift galaxies necessary.
      
It is impossible to observe of Pop III stars during their lifetime at
high redshift, but they might still possibly be detected directly by
looking for their PISN explosions before their death.  This idea has
been studied and shown to be promising for both LSST \citep{Trenti09}
and JWST \citep{Hummel12,Whalen13}. Understanding the population and
distribution of Pop III stars and remnants in high redshift galaxies
is helpful for preparing the observation of these events.

In this paper, we focus on the the formation of Pop III stars and the
population and multiplicity of Pop III stars and remnants in high
redshift galaxies. We have performed a simulation of a survey volume
of over 100 comoving Mpc$^{3}$ that includes a full primordial
chemistry network, radiative cooling from metal species, both Pop II and Pop
III star formation and their radiative, thermal, mechanical and
chemical feedback. The simulation runs on such a large volume from
cosmological initial conditions, so that we can follow the formation and
fate of Pop III stars in a statistically significant number of halos
with a wide range of masses from a few million solar mass to one
billion solar mass.  We first describe our simulation model in Section
2. Then in Section 3, we present our results of the Pop III stars and
remnants distribution over early galaxies. We discuss the findings and
possible bias in our simulation in Section 4.

\section{Simulation Setup}
\label{simulation}

We perform the simulation using the adaptive mesh refinement (AMR)
cosmological hydrodynamics code Enzo \citep{OShea04}.  Adaptive ray
tracing \citep{Wise11} is used for the radiation transfer of ionizing
radiation, which is coupled to the hydrodynamics and chemistry in
Enzo. The chemistry, cooling, and star formation and feedback models
used in this simulation are the same as in \citet{Wise12a}.

We generate the initial conditions for the simulation using
\textsc{Music} \citep{Hahn11_MUSIC} at $z=99$ and use the cosmological
parameters from the 7-year WMAP $\Lambda$CDM+SZ+LENS best fit
\citep{Komatsu11}: $\Omega_{M}$=0.266, $\Omega_{\Lambda}$ = 0.734,
$\Omega_{b}$=0.0449, h=0.71, $\sigma_{8}$=0.81, and n=0.963.  We use a
comoving simulation box of (40 Mpc)$^3$ that has a $512^3$ root grid
resolution and three levels of static nested grids.  We first run a
512$^3$ N-body only simulation to $z=6$. Then we select the Lagrangian
volume around two $\sim 3 \times 10^{10} \Ms$ halos at $z=6$, and
re-initialize the simulation with 3 more nested grids to have an
effective resolution of 4096$^3$ and an effective dark matter mass
resolution of $2.9 \times 10^4 \Ms$ inside the highest nested grid
with a comoving volume of $5.2 \times 7.0 \times 8.3\,
  \mathrm{Mpc}^3$ (300 Mpc$^3$).  During the course of the
simulation, we allow a maximum refinement level $l=12$, resulting in a
maximal resolution of 19 comoving pc.  The refinement
criteria employed are also the same as in \citet{Wise12a}.
The refinements higher than the static nested grids are only
  allowed in the Lagrangian volume of the two massive halos at $z=6$,
  which contains only high resolution particles.  It has a comoving
  volume of $3.8 \times 5.4 \times 6.6$ Mpc$^3$ ($\sim$138 Mpc$^3$) at
  $z=15$.  This highly refined volume is also the survey volume of
  this study. At this time, the simulation has a large number
($\sim$1000) of halos with M $>$ 10$^8$ M$_{\odot}$, where new
formation of Pop III stars declines rapidly, for statistical
analysis. We use results at this redshift for our current study.  The
simulation has 1.3 billion computational cells and required more than
10 million CPU hours, and there are three $>$ 10$^{9}$ M$_{\odot}$
halos in the refined regions at this time. We will continue this
simulation to lower redshift for the study of the Pop II stars and
high redshift galaxies.

Both Pop II and Pop III stars form in the simulation, and we
distinguish them by the total metallicity of the densest star forming
cell. Pop III stars are formed if $[\textrm{Z/H}] > -4$, and Pop II
stars are formed otherwise. We use the same star formation models and
most of the parameters in \citet{Wise12a}, as well as feedback
models. For the initial mass of Pop III stars, we randomly 
sample from an IMF with a functional form
\begin{equation}
f(\log M)dM=M^{-1.3}\exp\left[-\left(\frac{M_{\rm
        char}}{M}\right)^{1.6}\right]\, dM
\end{equation}
which behaves as a Salpeter IMF above the characteristic mass,
  M$_{\rm char}$, but is exponentially cutoff below that mass
  \citep{Chabrier03,Clark09}.  The only difference between the two
simulations is that we here use a characteristic mass of 40 M$_\odot$
for the Pop III IMF, which is more in line with the latest results of
Pop III formation simulations \citep[e.g.][]{Turk09,
  Greif12_P3Cluster}, instead of 100 M$_\odot$.  Please see Section
2.2 and Section 2.3 of \citet{Wise12a}, for the details of the star
formation and stellar feedback used in the simulation, respectively.

We show the evolution of the high-resolution region in Figure 1 at
three redshifts, 25, 17.91, and 15.  Here we show the large scale
structure in the inner 6.6 comoving Mpc in the top row, and in the
remaining rows, we focus on the most massive halo at each redshift by
showing their density-weighted projections of gas density,
temperature, and metallicity.

\section{Results}
\label{results}

\begin{figure}
\begin{center}
\epsfig{file=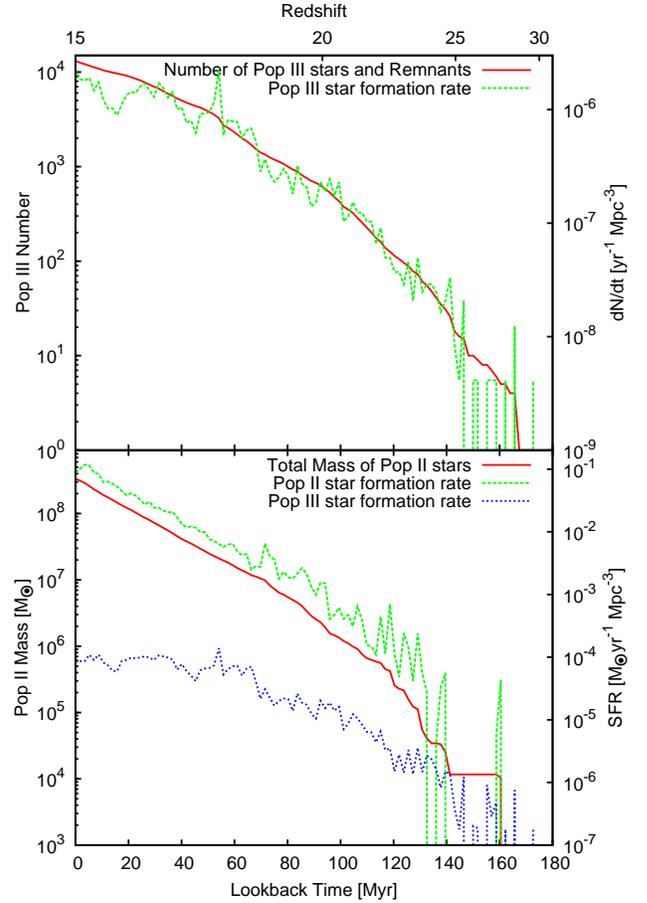,width=1.0\columnwidth}
\end{center}
\caption{Top panel: Evolution of the cumulative number of Population III stars and remnants and the averaged  
Pop III star number formation rate inside the survey volume. Bottom panel: Evolution of the total Pop 
II star mass, and the Pop II and Pop III star formation rate by mass. The mass formation rate of Pop III stars depends
on the choice of M$_{char}$.
\label{fig:evolution_PopIII}}
\end{figure} 

\begin{figure*}
\begin{center}
\epsfig{file=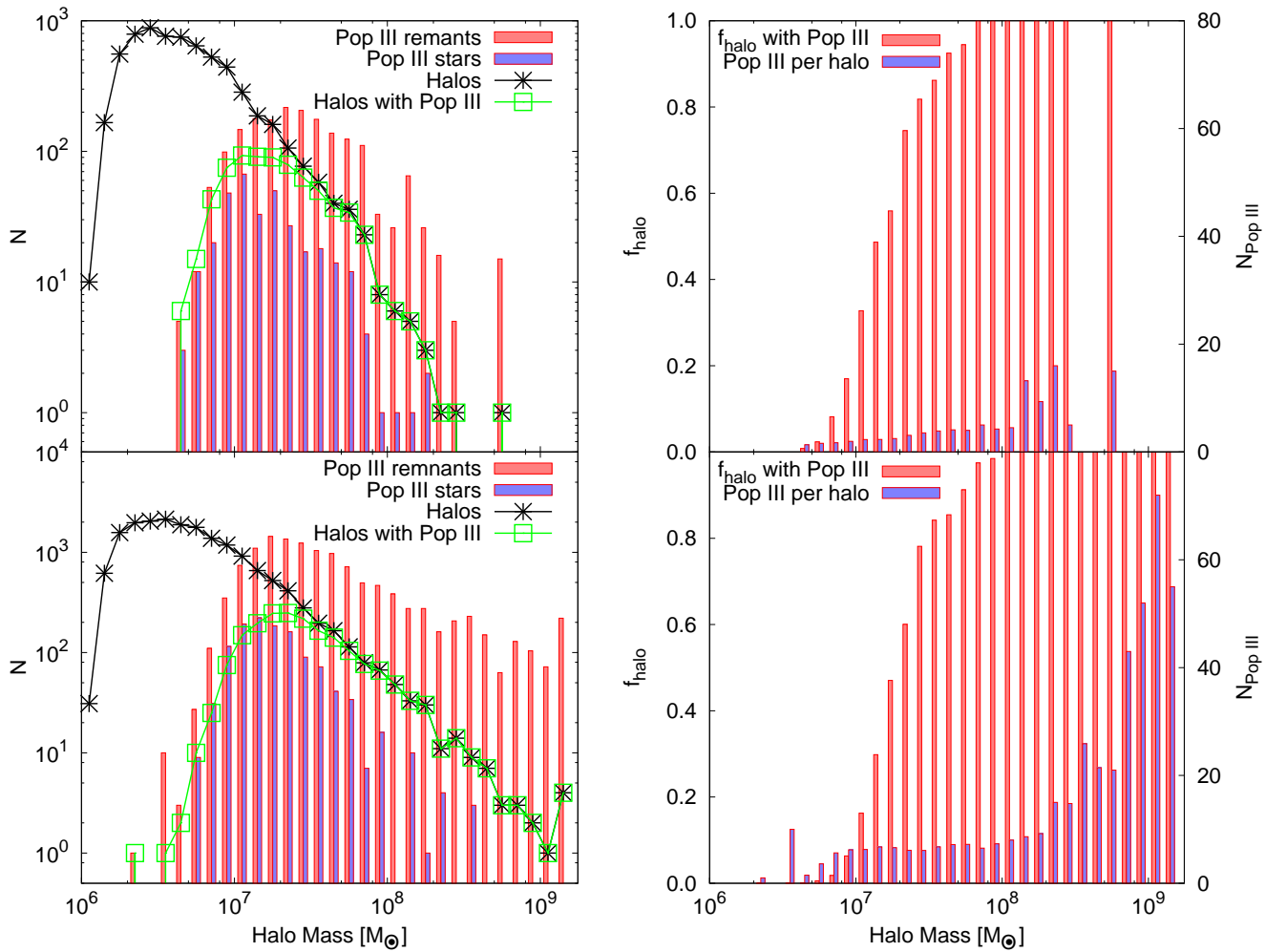,width=1.0\textwidth}
\end{center}
\caption{Left panel: Number of Pop III, total halos and halos with Pop III as functions of halo 
mass at redshifts 17.91 (Top) and 15 (Bottom). The number of living Pop III stars 
and the Pop III remnants are shown in different boxes. 
Right panel: Fraction of halos hosting Pop III and average number of Pop III
per halo at redshift 17.91 (Top) and 15 (Bottom). 
\label{fig:Histogram_PopIII_Halo}}
\end{figure*} 

We illustrate the evolution of the number of Pop III stars and
remnants, as well as their formation rate from the birth of the first
Pop III star to $z=15$ in the top panel of Figure 2. Massive Pop III
stars have very short lifetimes \citep{Schaerer02} and end their life
by either directly collapsing to black holes or exploding as
supernovae \citep{Heger03}, depending on their initial masses. 
More specifically, they die as Type II SNe if 11 $\le$ M$_\star$ / M$_\odot$ 
$\le$ 40 and as PISNe if 140 $\le$ M$_\star$ / M$_\odot$ $\le$ 260, where M$_\star$
is the initial stellar mass, or become black holes if their masses are 
not in these mass ranges.  
Throughout the paper, we use ``Pop III remnants'' to refer to all
remains of dead Pop III stars, regardless whether they become black
holes or supernova remnants, and we use ``Pop III'' to refer to {\it
  both} living stars and dead remnants.  In the case that remnants
have negligible masses after SN events, their star masses are replaced
with a very small mass proportional to their initial masses and the
star particles are kept in the simulation to follow their remnant
kinematic distribution.  The first Pop III star forms at redshift $z
\sim 29.7$. Rates of Pop III star formation steadily increase from
$\sim$ 10$^{-8}$ to above 10$^{-6}$ stars per year per comoving
Mpc$^3$. The Pop III star formation rate shows some signs of
saturation at $z \sim 15$ in this overdense region of the universe.
At $z=15$, the entire survey volume of 138 comoving Mpc$^3$ contains
13,123 Pop III stars and remnants and 7,677 halos more massive than 5
$\times$ 10$^6$ M$_{\odot}$.  The number densities of Pop III and
halos that host Pop III are 95 and 55 per comoving Mpc$^3$,
respectively.

\begin{figure}
\begin{center}
\epsfig{file=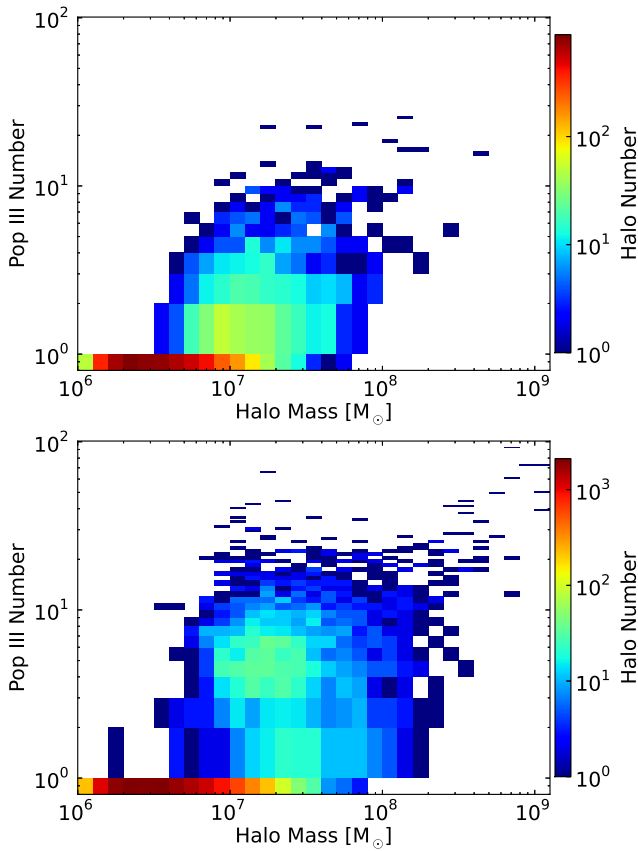,width=1.0\columnwidth}
\end{center}
\caption{ Phase plots of halo number indicating the distribution of
  Pop III stars and remnants over halos as a function of Pop III number and their host halo mass at redshifts 17.91 (Top) and 15 (Bottom). 
  \label{fig:Phase_Halo_PopIII}}
\end{figure} 

\begin{figure}
\begin{center}
\epsfig{file=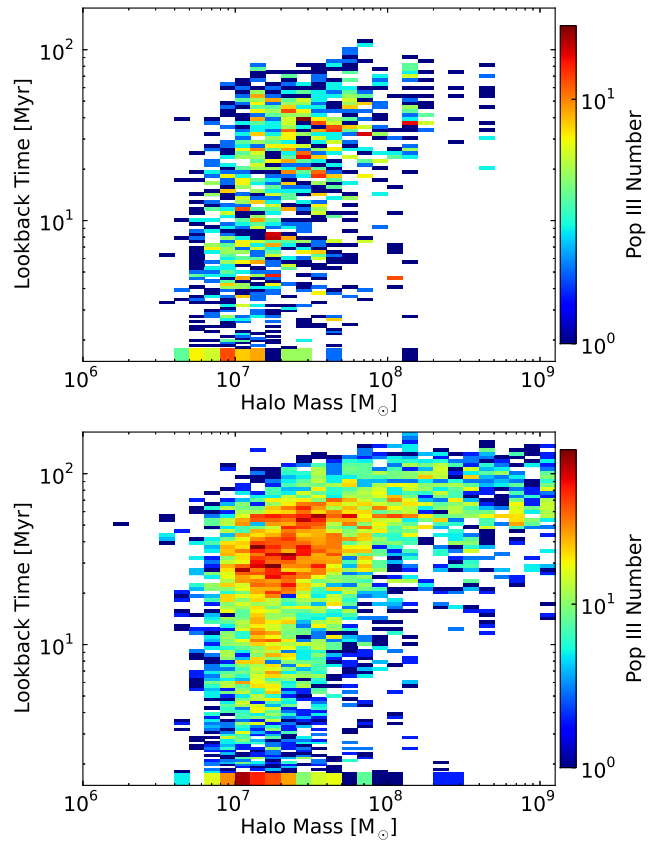,width=1.0\columnwidth}
\end{center}
\caption{Phase plots of Pop III number indicating the relation between
  Pop III formation and host halo mass as a function of Pop III star formation time and 
their host halo mass at redshifts 17.91 (Top) and 15 (Bottom). 
\label{fig:Phase_Halo_PopIIItime}}
\end{figure}

The total mass of Pop II stars and the SFR of Pop II and Pop III are
shown in the bottom panel of Figure 2. There is $\sim$ 3 $\times$
10$^{8}$ M$_{\odot}$ mass in Pop II stars at $z=15$.  The star formation
histories for both Pop II and Pop III stars are similar to those in
\citet{Wise12a}, but shifted to higher redshifts.  We will study the
details of Pop II star formation in this simulation in a forthcoming
paper.

We show the number of Pop III, all halos, and star-hosting halos as
function of halo mass inside the survey volume at $z=17.91$ and $z=15$
in the left panels in Figure 3, while the numbers of living Pop III
stars are stacked over remnants in the same bin.  In the right panels
in Figure 3, shown are the fraction of halos hosting Pop III and the
average Pop III per halo (over halos having Pop III) as functions of
halo mass.

As our survey volume is chosen over a high density region, the halo
number density is higher than the Universe mean. The halo mass
functions from our simulation are well fit with the fitting function
of \citet{Warren06} boosted by a factor of five for halos heavier than
5 $\times$ 10$^{6}$ M$_{\odot}$ in both redshifts.  The number of
halos drops for halos $<$ 5 $\times$ 10$^{6}$ M$_{\odot}$, due to the
lack of dark matter mass resolution. At $z=15$, our simulation has a
factor of a few more halos than the prediction from the fitting
function for halos more massive than 1 $\times$ 10$^8$ $M_{\odot}$, showing that
the baryon becomes important in massive halos. There are 3 halos with
masses over 10$^{9}$ M$_\odot$ at $z=15$, while the first one appears
at redshift $z \sim 15.8$.

The Pop III distribution over halo mass shows little evolution, except
that there are more Pop III appearing in high mass halos as those
halos building up with time.  The number of Pop III peaks at $\sim$ 3
$\times$ 10$^7$ M$_{\odot}$. The fraction of halos hosting Pop III
also has little change between redshifts 17.91 and 15 because the
chances to host Pop III are very small for halos less massive than
10$^7$ M$_\odot$ in this biased region.  Then they gradually increases
with halo mass and reach 100\% for halos $\sim$ 10$^8$ M$_\odot$.  In
this particular mass range of $10^7 - 10^8 \Ms$, halos can cool and
collapse regardless of the LW background and form stars
\citep[e.g.][]{Wise07_UVB, OShea08, Shang10, Wolcott11}.  The exact
timing of Pop III star formation depends on the local strength of the
LW intensity, causing the gradual increase in the fraction shown in
Figure 3.  In addition to \hh~suppression, nearby galaxies and Pop III
supernova can chemical enrich halos so that they form metal-enriched
stars in their first star formation event without hosting any prior
Pop III star formation.  Afterwards, as halos grow through hierarchical
mergers, the number of Pop III per halos increase with both redshift
and halo mass.  At $z=15$, halos between 10$^7$ M$_\odot$ and 10$^8$
M$_\odot$ have an average of 10 Pop III, and the number rises to about
50 for halos $\sim$ 10$^9$ M$_\odot$.

Shown in Figure 4 are the number distribution of halos as a function
of halo mass and number of Pop III stars and remnants per halo.  The
number of Pop III that is hosted by halos inside a mass bin is very
scattered, but there is a clear trend that more massive halos host
more Pop III stars and remnants. The same trend is shown in the
average number of Pop III per halo in Figure 3. At redshift $z=15$,
the most massive halo without any Pop III stars or remnants is 7.16
$\times$ 10$^{7}$ M$_{\odot}$. On the low mass end, the least massive
halo hosting Pop III is only 1.82 $\times$ 10$^{6}$ M$_{\odot}$. But
this Pop III star forms at about 50 Myr ago, so this star most likely
forms in another more massive halo, then is stripped out with its
current host halo from the massive one.

In Figure 5, we show the number distribution of Pop III as function of
their host halo mass and their formation time.  As old Pop III
remnants can be found in halos of all masses, young stars and remnants
can only be seen in low mass halos.  No Pop III star forms in halos
more massive than 3 $\times$ 10$^{8}$ M$_{\odot}$. The gas in massive
halos is enriched by their earlier Pop III, so the further formation
of stars transitions to Pop II. The least massive halos having newly
formed Pop III stars is $\sim$ 4 $\times$ 10$^6$ M$_{\odot}$, as the
Pop III formation in smaller halos is suppressed by the LW
background. But our results for Pop III formation in low mass halos
might be biased. Due to the relatively low dark matter (2.9 $\times$
10$^4$ M$_{\odot}$), we may miss the formation of $M \lsim 10^6 \Ms$
halos in our simulation. We discuss this issue in the next section and
show that very little Pop III star formation is missed by the current
simulation due to resolution effects.

\section{Discussions and Conclusions}
\label{summary}

\begin{figure}[b]
\begin{center}
\epsfig{file=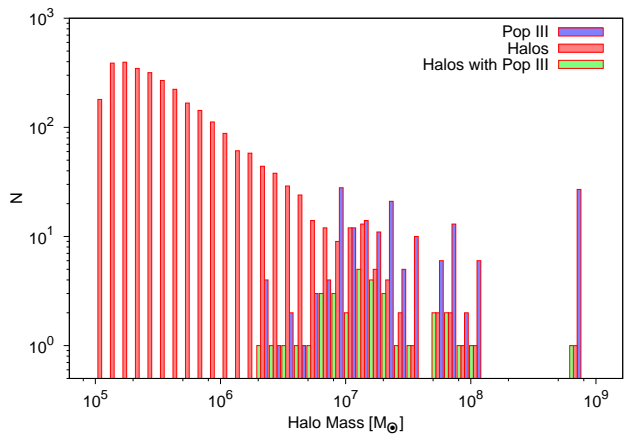,width=1.0\columnwidth}
\end{center}
\caption{Number of Pop III stars and remnants, all halos, and halos
  with Pop III as functions of halo mass of the 1 comoving Mpc$^3$
  simulation in \citet{Wise12a} at redshift 7.28. With a dark matter
  mass resolution that is finer by a factor of 16, the halo population
  extends from 10$^6$ M$_{\odot}$ to 2 $\times$ 10$^5$
  M$_{\odot}$. However, there are no Pop III present in those small
  halos, which are suppressed by a LW background.  Indeed, the
  distribution of Pop III is similar to our current results shown in
  Figure 3, though two survey volumes are in very different density
  fluctuations. \label{fig:Histogram_PopIII_Halo_1mpc}}
\end{figure} 

We present results of a cosmological simulation representing a high
density region of an unprecedented large volume in the early universe
to study the Pop III star and remnant population and multiplicity within
high redshift galaxies. The highly resolved simulation volume covers
thousand of halos and Pop III stars and remnants, allowing us to make a
detailed study of the Pop III star formation history and the Pop III
stars and remnants distribution over a wide range of halo masses. We
observe a continuous Pop III star formation from z $\sim$ 30 to z
$\sim$ 15, when the calculation was paused for analysis. We expect 
Pop III star formation to continue to lower redshifts, but this needs 
to be verified by running the simulation further. We find that the 
number of the Pop III stars and remnants peaks in halos with masses of 
a few $\times$ 10$^{7}$ M$_{\odot}$ during this time, while
the ratio to host Pop III and multiplicity of Pop III increase with
the halo mass. The Pop III inside a massive halos are likely coming
from mergers of small ones, instead of forming in the big ones. New
Pop III stars only form in halos between 4 $\times$ 10$^6$ to 3
$\times$ 10$^8$ M$_{\odot}$, while the formation rate peaks at $\sim$
10$^7$ M$_\odot$. So even though there are tens of Pop III remnants in a halo
of 10$^{9}$ M$_{\odot}$, little to no feedback might be expected
coming from their still living Pop III stars.

\begin{figure}[b]
\begin{center}
\epsfig{file=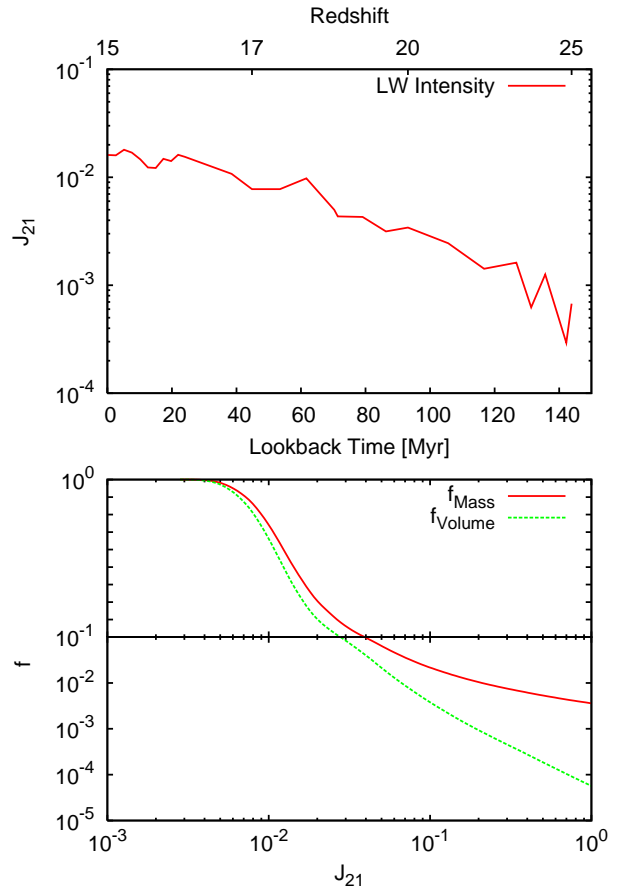,width=1.0\columnwidth}
\end{center}
\caption{Top panel: Evolution of the volume averaged Lyman-Werner
  radiation intensity from stars in the simulation.  Bottom panel:
  Complementary cumulative distributions of the LW intensity at
  $z=15$. J$_{21}$ is the LW intensity in the unit of 10$^{-21}$ erg
  s$^{-1}$ cm$^{-2}$ Hz$^{-1}$ sr$^{-1}$.  The averaged LW intensity
  gradually increases to and maintains above 10$^{-23}$ erg s$^{-1}$
  cm$^{-2}$ Hz$^{-1}$ sr$^{-1}$ for the last 40 Myr. This LW radiation
  is not enough to significantly slow down the Pop III formation.
\label{fig:LW_evolution}}
\end{figure}

The dark matter mass resolution in this simulation is not high enough
to capture all of the Pop III star forming halos smaller that a few
$\times$10$^{6}$ M$_\odot$, so we could be underestimating the number
of Pop III in our simulation. To investigate this, we use the simulation in
\citet{Wise12b} to estimate the missing Pop III stars and remnants
in low mass halos.  We show the histogram of halo mass in that
1 comoving Mpc$^3$ box simulation at $z = 7.28$ in Figure 6. Due to the
much higher dark matter mass resolution, there are many more halos at
a few 10$^6$ M$_{\odot}$, extending to $\sim$ 2 $\times$ 10$^{5}$
M$_{\odot}$.  Since the simulation box is small, the halo mass
function is not well fit by the \citet{Warren06} function by
having too many low mass halos and too few high mass halos, but still
has the number of halos $\sim$ 10$^{6}$ M$_{\odot}$ close to cosmic
mean. Though there are much less halos and Pop III in this smaller
simulated volume, it has a similar distribution of Pop III stars and
remnants as in our current simulation. Halos with $M_{\rm halo} \la 2
\times 10^{6} \Ms$ do not host any Pop III stars or remnants in this
small box simulation.  Pop III star formation in these low-mass halos
is suppressed when the \hh~is dissociated by LW radiation from nearby
stellar sources and the background \citep{Machacek01, Wise07_UVB,
  OShea08}.  The first star forms at $z \sim 20$, when the LW
background is already at an intensity of $\sim 10^{-23}$ \emis, and
suppresses any \hh~formation and thus Pop III star formation in halos
below a mass of $10^6 \Ms$.  At later times, we also confirm that the
minimum star-forming halo mass with this LW background in
\citet{Wise12b} at any time is $\ga 10^6 \Ms$.  
Thus, we conclude that the Pop III formation rate is small in halos 
below this mass scale and that we do not miss a significant fraction of 
Pop III star formation in our simulation because of mass resolution.

In this large volume simulation, only the LW flux from point sources
is included, while the LW background is ignored. We show the evolution
of averaged LW intensity and the fraction distribution of LW intensity
at $z=15$ over the refined volume in Figure \ref{fig:LW_evolution}.
The averaged LW radiation intensity builds up gradually and reaches
10$^{-23}$ \emis at redshift $z \sim 17$, and maintains at this level
for more than 40 Myr. At $z=15$, the LW intensity in $\sim 70\%$ of
the volume is higher than 10$^{-23}$ \emis , while only less than
0.01\% of the volume (0.014 comoving (Mpc)$^3$) is filled with LW
radiation stronger than 10$^{-21}$ \emis. This LW intensity, locally
produced even in a rare high density region, is still weaker than the
expected LW background \citep{Wise05, Wise12b}. Usually, it is
  thought that this level of LW radiation may only suppress Pop III
  formation in $\sim$10$^6$ $M_\odot$ halos but is not strong enough
  to have a significant impact in larger minihalos ($3 \times 10^6 -
  10^7 M_\odot$).  However, we found that only 20\% of halos in this
  mass range form Pop III stars, delayed by LW feedback.  At $z=15$,
  90 percent (two-thirds) of the halos with no Pop III stars and $M =
  5 \times 10^6 - 3 \times 10^7 M_\odot$ have LW intensities higher
  than 10$^{-23}$ (2 $\times$ 10$^{-23}$) \emis.  The main difference
  in our simulation to previous studies of LW feedback is environment.
  In our simulated overdense region, the halo mass accretion rates are
  high enough so that dynamical heating \citep{Yoshida03} becomes
  important, where mergers and rapid cosmological inflows cause
  shock-heating and virial turbulence \citep{Wise07, Greif08} that can
  disrupt any dense central cores.  Embedding these rapidly growing
  minihalos in a LW background delays their collapse even further as
  these disrupted cores must cool once again and re-coalesce.  The
  typical collapse time to form Pop III stars in a 10$^{-23}$ \emis\
  LW background is 50~Myr \citep{OShea08}.  Most of the minihalos in
  our simulation at these very high redshifts are younger than 50~Myr,
  and thus have not had sufficient time to collapse.

An important difference between the simulation in this paper and that of
\citet{Wise12b} is that the characteristic mass of Pop III stars here
is much lower at M$_{char}$ = 40 M$_{\odot}$, instead of M$_{char}$ =
100 M$_{\odot}$. The new choice of characteristic mass results that
much fewer Pop III stars have initial mass between 140 and 260
M$_\odot$, and end their life by PISN. This then significantly reduces
the metal generated by Pop III stars. For example, a 40 M$_\odot$
hypernova produces only 8.6 M$_\odot$ metals \citep{Nomoto06},
comparing to 85 M$_\odot$ of metals generated by 180 M$_\odot$ PISN
\citep{Heger02}. It is likely due to the lower level of metal
generation, that Pop III stars continue to form in the larger halos ($>$ 10$^8$
M$_\odot$) in current simulation.  On the contrary, there is no Pop
III formed in halos more massive than 5 $\times$ 10$^7$ M$_{\odot}$ in
\citet{Wise12a}.

\begin{figure}[t]
\begin{center}
\epsfig{file=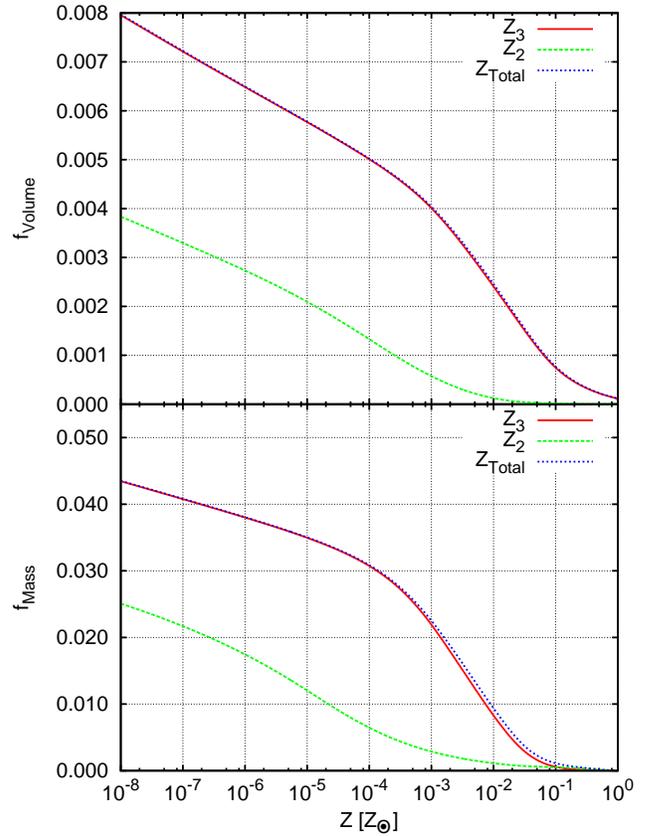,width=1.0\columnwidth}
\end{center}
\caption{Complementary cumulative distributions of the metallicity by volume (Top) and by
mass (Bottom) inside the refined region at $z=15$. The solid, dashed, dotted lines show the metal from 
Pop III stars, Pop II stars, and the total metal, respectively. The total metallicity 
distributions are still determined by the feedback from Pop III stars at this epoch.  
\label{fig:Metallicity_distribution}}
\end{figure}

Because our simulation covers a rare high density region, our results
are not representative of the cosmic mean.  We fit the halo mass
function with the fitting function of \citet{Warren06}. It
shows that the halo number density at $z=15$ is about five times that of
the cosmic mean at the same redshift or about the cosmic mean at $z=10$,
while the halo number density at $z = 17.91$ is about the cosmic mean at
$z = 14$. The density fluctuations may have effects on local Pop III
formation by changing the LW background and metallicity level of the
IGM.  As discussed before, the strength of LW background during these
redshifts might not change the Pop III formation significantly.  In
addition, because the mean distance between Pop III and halos is still
much larger than the direct metal impact distance of a PISN of 5 kpc
\citep{Wise12a}, the metal enrichment by Pop III feedback crossing
halos is still weak, except in halo clusters. In the case, the Pop III
formation mostly depends on single halo properties, while the
multiplicity of Pop III over halos is determined by the clustering of
halos. Our results of Pop III distribution and multiplicity over halos
should not change significantly in regions of different density
fluctuations. The population of Pop III is proportional to the halo
density, so the number density from this simulation at $z=15$ should be
about 5 times that of the cosmic mean or close to the cosmic mean at z
$\sim$ 10.

One important question about Pop III stars is when and in what environments 
their formation completely ends. Unfortunately,  our simulation does 
not show the cessation, or even the saturation of Pop III formation yet,
though the increase of star formation rate slows down in the last tens of
million years. The Pop III star and remnant number density is
still low (lower than that of \citet{Wise12a} 1 Mpc$^{3}$ simulation),
and there still is a significant fraction of halos between 10$^7$ and
10$^8$ M$_\odot$ that have no Pop III stars and are in a favorable LW 
radiation field. To show the sufficiency of metal poor gas inside the survey 
volume to continue form Pop III stars, we plot the volume and mass 
complementary cumulative distributions 
of the metallicity at $z=15$ in Figure \ref{fig:Metallicity_distribution}. 
There are only about 0.65\% of volume and 3.8\% of the mass enriched 
above 10$^{-6}$ Z$_\odot$. For comparison, the \citet{Wise12a} 1 Mpc$^{3}$ 
simulation has 0.32\% of the volume and 2.6\% of the mass at $z=10$, 
and 0.89\% of the volume and 7.6\% of the mass at $z=7.28$,
respectively, above 10$^{-6}$ Z$_\odot$.  So it is reasonable to
expect Pop III stars will continue to form inside our survey volume,
but at a constant rate, since \citet{Wise12a} shows that the Pop III
formation rate is constant from $z = 12$ to 7. We are continuing the 
present simulation to lower redshifts to confirm or deny this expectation.

\acknowledgements 

We thank an anonymous referee for helpful comments. This
research was supported by National Science Foundation (NSF) grant
AST-1109243 to MLN.  JHW acknowledges support from NSF grant
AST-1211626.  The simulation was performed on the Kraken supercomputer
operated for the Extreme Science and Engineering Discovery Environment
(XSEDE) by the National Institute for Computational Science, ORNL with
XRAC allocation MCA-TG98020N. This research has made use of
  NASA's Astrophysics Data System Bibliographic Services.  The
  majority of the analysis and plots were done with \texttt{yt}
  \citep{yt_full_paper}.


\begin{thebibliography}{55}
\expandafter\ifx\csname natexlab\endcsname\relax\def\natexlab#1{#1}\fi

\bibitem[{{Abel} {et~al.}(2002){Abel}, {Bryan}, \& {Norman}}]{Abel02}
{Abel}, T., {Bryan}, G.~L., \& {Norman}, M.~L. 2002, Science, 295, 93

\bibitem[{{Alvarez} {et~al.}(2009){Alvarez}, {Wise}, \& {Abel}}]{Alvarez09}
{Alvarez}, M.~A., {Wise}, J.~H., \& {Abel}, T. 2009, \apjl, 701, L133

\bibitem[{{Bromm} {et~al.}(2002){Bromm}, {Coppi}, \& {Larson}}]{Bromm02}
{Bromm}, V., {Coppi}, P.~S., \& {Larson}, R.~B. 2002, \apj, 564, 23

\bibitem[{{Bromm} {et~al.}(2001){Bromm}, {Ferrara}, {Coppi}, \&
  {Larson}}]{Bromm01}
{Bromm}, V., {Ferrara}, A., {Coppi}, P.~S., \& {Larson}, R.~B. 2001, \mnras,
  328, 969

\bibitem[{{Chabrier}(2003)}]{Chabrier03}
{Chabrier}, G. 2003, \pasp, 115, 763

\bibitem[{{Clark} {et~al.}(2009){Clark}, {Glover}, {Bonnell}, \&
  {Klessen}}]{Clark09}
{Clark}, P.~C., {Glover}, S.~C.~O., {Bonnell}, I.~A., \& {Klessen}, R.~S. 2009,
  ArXiv e-prints (0904.3302)

\bibitem[{{Clark} {et~al.}(2008){Clark}, {Glover}, \& {Klessen}}]{Clark08}
{Clark}, P.~C., {Glover}, S.~C.~O., \& {Klessen}, R.~S. 2008, \apj, 672, 757

\bibitem[{{Greif} {et~al.}(2012){Greif}, {Bromm}, {Clark}, {Glover}, {Smith},
  {Klessen}, {Yoshida}, \& {Springel}}]{Greif12_P3Cluster}
{Greif}, T.~H., {Bromm}, V., {Clark}, P.~C., {Glover}, S.~C.~O., {Smith},
  R.~J., {Klessen}, R.~S., {Yoshida}, N., \& {Springel}, V. 2012, \mnras, 424,
  399

\bibitem[{{Greif} {et~al.}(2008){Greif}, {Johnson}, {Klessen}, \&
  {Bromm}}]{Greif08}
{Greif}, T.~H., {Johnson}, J.~L., {Klessen}, R.~S., \& {Bromm}, V. 2008,
  \mnras, 387, 1021

\bibitem[{{Hahn} \& {Abel}(2011)}]{Hahn11_MUSIC}
{Hahn}, O., \& {Abel}, T. 2011, \mnras, 415, 2101

\bibitem[{{Haiman} {et~al.}(2000){Haiman}, {Abel}, \& {Rees}}]{Haiman00}
{Haiman}, Z., {Abel}, T., \& {Rees}, M.~J. 2000, \apj, 534, 11

\bibitem[{{Heger} {et~al.}(2003){Heger}, {Fryer}, {Woosley}, {Langer}, \&
  {Hartmann}}]{Heger03}
{Heger}, A., {Fryer}, C.~L., {Woosley}, S.~E., {Langer}, N., \& {Hartmann},
  D.~H. 2003, \apj, 591, 288

\bibitem[{{Heger} \& {Woosley}(2002)}]{Heger02}
{Heger}, A., \& {Woosley}, S.~E. 2002, \apj, 567, 532

\bibitem[{{Hummel} {et~al.}(2012){Hummel}, {Pawlik}, {Milosavljevi{\'c}}, \&
  {Bromm}}]{Hummel12}
{Hummel}, J.~A., {Pawlik}, A.~H., {Milosavljevi{\'c}}, M., \& {Bromm}, V. 2012,
  \apj, 755, 72

\bibitem[{{Johnson} {et~al.}(2012){Johnson}, {Whalen}, {Li}, \&
  {Holz}}]{Johnson12}
{Johnson}, J.~L., {Whalen}, D.~J., {Li}, H., \& {Holz}, D.~E. 2012, ArXiv
  e-prints

\bibitem[{{Karlsson} {et~al.}(2008){Karlsson}, {Johnson}, \&
  {Bromm}}]{Karlsson08}
{Karlsson}, T., {Johnson}, J.~L., \& {Bromm}, V. 2008, \apj, 679, 6

\bibitem[{{Komatsu} {et~al.}(2011){Komatsu}, {Smith}, {Dunkley}, {Bennett},
  {Gold}, {Hinshaw}, {Jarosik}, {Larson}, {Nolta}, {Page}, {Spergel},
  {Halpern}, {Hill}, {Kogut}, {Limon}, {Meyer}, {Odegard}, {Tucker}, {Weiland},
  {Wollack}, \& {Wright}}]{Komatsu11}
{Komatsu}, E., {Smith}, K.~M., {Dunkley}, J., {Bennett}, C.~L., {Gold}, B.,
  {Hinshaw}, G., {Jarosik}, N., {Larson}, D., {Nolta}, M.~R., {Page}, L.,
  {Spergel}, D.~N., {Halpern}, M., {Hill}, R.~S., {Kogut}, A., {Limon}, M.,
  {Meyer}, S.~S., {Odegard}, N., {Tucker}, G.~S., {Weiland}, J.~L., {Wollack},
  E., \& {Wright}, E.~L. 2011, \apjs, 192, 18

\bibitem[{{Komiya} {et~al.}(2010){Komiya}, {Habe}, {Suda}, \&
  {Fujimoto}}]{Komiya10}
{Komiya}, Y., {Habe}, A., {Suda}, T., \& {Fujimoto}, M.~Y. 2010, \apj, 717, 542

\bibitem[{{Larson}(2005)}]{Larson05}
{Larson}, R.~B. 2005, \mnras, 359, 211

\bibitem[{{Machacek} {et~al.}(2001){Machacek}, {Bryan}, \& {Abel}}]{Machacek01}
{Machacek}, M.~E., {Bryan}, G.~L., \& {Abel}, T. 2001, \apj, 548, 509

\bibitem[{{Maio} {et~al.}(2010){Maio}, {Ciardi}, {Dolag}, {Tornatore}, \&
  {Khochfar}}]{Maio10_Pop32}
{Maio}, U., {Ciardi}, B., {Dolag}, K., {Tornatore}, L., \& {Khochfar}, S. 2010,
  \mnras, 407, 1003

\bibitem[{{Mesinger} {et~al.}(2013){Mesinger}, {Ferrara}, \&
  {Spiegel}}]{Mesinger13}
{Mesinger}, A., {Ferrara}, A., \& {Spiegel}, D.~S. 2013, \mnras

\bibitem[{{Muratov} {et~al.}(2012){Muratov}, {Gnedin}, {Gnedin}, \&
  {Zemp}}]{Muratov12}
{Muratov}, A.~L., {Gnedin}, O.~Y., {Gnedin}, N.~Y., \& {Zemp}, M. 2012, ArXiv
  e-prints

\bibitem[{{Nomoto} {et~al.}(2006){Nomoto}, {Tominaga}, {Umeda}, {Kobayashi}, \&
  {Maeda}}]{Nomoto06}
{Nomoto}, K., {Tominaga}, N., {Umeda}, H., {Kobayashi}, C., \& {Maeda}, K.
  2006, Nuclear Physics A, 777, 424

\bibitem[{{Omukai} {et~al.}(2005){Omukai}, {Tsuribe}, {Schneider}, \&
  {Ferrara}}]{Omukai05}
{Omukai}, K., {Tsuribe}, T., {Schneider}, R., \& {Ferrara}, A. 2005, \apj, 626,
  627

\bibitem[{{O'Shea} {et~al.}(2004){O'Shea}, {Bryan}, {Bordner}, {Norman},
  {Abel}, {Harkness}, \& {Kritsuk}}]{OShea04}
{O'Shea}, B.~W., {Bryan}, G., {Bordner}, J., {Norman}, M.~L., {Abel}, T.,
  {Harkness}, R., \& {Kritsuk}, A. 2004, ArXiv Astrophysics e-prints
  (astro-ph/0403044)

\bibitem[{{O'Shea} \& {Norman}(2007)}]{OShea07a}
{O'Shea}, B.~W., \& {Norman}, M.~L. 2007, \apj, 654, 66

\bibitem[{{O'Shea} \& {Norman}(2008)}]{OShea08}
---. 2008, \apj, 673, 14

\bibitem[{{Ricotti} {et~al.}(2008){Ricotti}, {Gnedin}, \& {Shull}}]{Ricotti08}
{Ricotti}, M., {Gnedin}, N.~Y., \& {Shull}, J.~M. 2008, \apj, 685, 21

\bibitem[{{Ricotti} \& {Ostriker}(2004)}]{Ricotti04}
{Ricotti}, M., \& {Ostriker}, J.~P. 2004, \mnras, 352, 547

\bibitem[{{Salvadori} {et~al.}(2007){Salvadori}, {Schneider}, \&
  {Ferrara}}]{Salvadori07}
{Salvadori}, S., {Schneider}, R., \& {Ferrara}, A. 2007, \mnras, 381, 647

\bibitem[{{Scannapieco} {et~al.}(2003){Scannapieco}, {Schneider}, \&
  {Ferrara}}]{Scannapieco03}
{Scannapieco}, E., {Schneider}, R., \& {Ferrara}, A. 2003, \apj, 589, 35

\bibitem[{{Schaerer}(2002)}]{Schaerer02}
{Schaerer}, D. 2002, \aap, 382, 28

\bibitem[{{Schneider} {et~al.}(2006){Schneider}, {Omukai}, {Inoue}, \&
  {Ferrara}}]{Schneider06}
{Schneider}, R., {Omukai}, K., {Inoue}, A.~K., \& {Ferrara}, A. 2006, \mnras,
  369, 1437

\bibitem[{{Shang} {et~al.}(2010){Shang}, {Bryan}, \& {Haiman}}]{Shang10}
{Shang}, C., {Bryan}, G.~L., \& {Haiman}, Z. 2010, \mnras, 402, 1249

\bibitem[{{Smith} {et~al.}(2009){Smith}, {Turk}, {Sigurdsson}, {O'Shea}, \&
  {Norman}}]{Smith09}
{Smith}, B.~D., {Turk}, M.~J., {Sigurdsson}, S., {O'Shea}, B.~W., \& {Norman},
  M.~L. 2009, \apj, 691, 441

\bibitem[{{Stacy} \& {Bromm}(2012)}]{Stacy12}
{Stacy}, A., \& {Bromm}, V. 2012, ArXiv e-prints

\bibitem[{{Stacy} {et~al.}(2010){Stacy}, {Greif}, \& {Bromm}}]{Stacy10}
{Stacy}, A., {Greif}, T.~H., \& {Bromm}, V. 2010, \mnras, 403, 45

\bibitem[{{Tornatore} {et~al.}(2007){Tornatore}, {Ferrara}, \&
  {Schneider}}]{Tornatore07}
{Tornatore}, L., {Ferrara}, A., \& {Schneider}, R. 2007, \mnras, 382, 945

\bibitem[{{Trenti} {et~al.}(2009){Trenti}, {Stiavelli}, \& {Shull}}]{Trenti09}
{Trenti}, M., {Stiavelli}, M., \& {Shull}, J.~M. 2009, \apj, 700, 1672

\bibitem[{{Tumlinson}(2006)}]{Tumlinson06}
{Tumlinson}, J. 2006, \apj, 641, 1

\bibitem[{{Turk} {et~al.}(2009){Turk}, {Abel}, \& {O'Shea}}]{Turk09}
{Turk}, M.~J., {Abel}, T., \& {O'Shea}, B. 2009, Science, 325, 601

\bibitem[{{Turk} {et~al.}(2011){Turk}, {Smith}, {Oishi}, {Skory}, {Skillman},
  {Abel}, \& {Norman}}]{yt_full_paper}
{Turk}, M.~J., {Smith}, B.~D., {Oishi}, J.~S., {Skory}, S., {Skillman}, S.~W.,
  {Abel}, T., \& {Norman}, M.~L. 2011, \apjs, 192, 9

\bibitem[{{Warren} {et~al.}(2006){Warren}, {Abazajian}, {Holz}, \&
  {Teodoro}}]{Warren06}
{Warren}, M.~S., {Abazajian}, K., {Holz}, D.~E., \& {Teodoro}, L. 2006, \apj,
  646, 881

\bibitem[{{Whalen} {et~al.}(2008){Whalen}, {O'Shea}, {Smidt}, \&
  {Norman}}]{Whalen08}
{Whalen}, D., {O'Shea}, B.~W., {Smidt}, J., \& {Norman}, M.~L. 2008, \apj, 679,
  925

\bibitem[{{Whalen} {et~al.}(2013){Whalen}, {Fryer}, {Holz}, {Heger}, {Woosley},
  {Stiavelli}, {Even}, \& {Frey}}]{Whalen13}
{Whalen}, D.~J., {Fryer}, C.~L., {Holz}, D.~E., {Heger}, A., {Woosley}, S.~E.,
  {Stiavelli}, M., {Even}, W., \& {Frey}, L.~H. 2013, \apjl, 762, L6

\bibitem[{{Wise} \& {Abel}(2005)}]{Wise05}
{Wise}, J.~H., \& {Abel}, T. 2005, \apj, 629, 615

\bibitem[{{Wise} \& {Abel}(2007{\natexlab{a}})}]{Wise07}
---. 2007{\natexlab{a}}, \apj, 665, 899

\bibitem[{{Wise} \& {Abel}(2007{\natexlab{b}})}]{Wise07_UVB}
---. 2007{\natexlab{b}}, \apj, 671, 1559

\bibitem[{{Wise} \& {Abel}(2011)}]{Wise11}
---. 2011, \mnras, 657

\bibitem[{{Wise} {et~al.}(2012{\natexlab{a}}){Wise}, {Abel}, {Turk}, {Norman},
  \& {Smith}}]{Wise12b}
{Wise}, J.~H., {Abel}, T., {Turk}, M.~J., {Norman}, M.~L., \& {Smith}, B.~D.
  2012{\natexlab{a}}, \mnras, 427, 311

\bibitem[{{Wise} {et~al.}(2012{\natexlab{b}}){Wise}, {Turk}, {Norman}, \&
  {Abel}}]{Wise12a}
{Wise}, J.~H., {Turk}, M.~J., {Norman}, M.~L., \& {Abel}, T.
  2012{\natexlab{b}}, \apj, 745, 50

\bibitem[{{Wolcott-Green} {et~al.}(2011){Wolcott-Green}, {Haiman}, \&
  {Bryan}}]{Wolcott11}
{Wolcott-Green}, J., {Haiman}, Z., \& {Bryan}, G.~L. 2011, \mnras, 418, 838

\bibitem[{{Yoshida} {et~al.}(2003){Yoshida}, {Abel}, {Hernquist}, \&
  {Sugiyama}}]{Yoshida03}
{Yoshida}, N., {Abel}, T., {Hernquist}, L., \& {Sugiyama}, N. 2003, \apj, 592,
  645

\bibitem[{{Yoshida} {et~al.}(2004){Yoshida}, {Bromm}, \&
  {Hernquist}}]{Yoshida04}
{Yoshida}, N., {Bromm}, V., \& {Hernquist}, L. 2004, \apj, 605, 579

\end{thebibliography}

\end{document}